# All-Fibre Label-Free Nano-Sensor for Real-Time in situ Early Monitoring of Cellular Apoptosis

Danran Li[1,2], Nina Wang[3], Tianyang Zhang[3], Guangxing Wu[1,2], Yifeng Xiong[1,2], Qianqian Du[4], Yunfei Tian[1,2], Wei-wei Zhao[3], Jiandong Ye[4], Shulin Gu[4], Yanqing Lu[1,2], Dechen Jiang[3,*] and Fei Xu[1,2,*]

[1]College of Engineering and Applied Sciences, Nanjing University, Nanjing 210093, China

[2]National Laboratory of Solid State Microstructures and Collaborative Innovation Center of Advanced Microstructures, Nanjing University, Nanjing 210093, China

[3]School of Chemistry and Chemical Engineering, Nanjing University, Nanjing 210093, China

[4]School of Electronic Science and Engineering, Nanjing University, Nanjing 210093, China

**\*Corresponding author**

Fei Xu        e-mail: feixu@nju.edu.cn

Dechen Jiang    e-mail: dechenjiang@nju.edu.cn

**Abstract**

The achievement of all-fibre functional nano-modules for subcellular label-free measurement has long been pursued due to the limitations of manufacturing techniques. In this paper, a compact all-fibre label-free nano-sensor composed of a fibre taper and zinc oxide nano-gratings is designed and applied for the early monitoring of apoptosis in single living cells. Because of its nanoscale dimensions, mechanical flexibility and minimal cytotoxicity to cells, the sensing module can be loaded in cells for long-term *in situ* tracking with high sensitivity. A gradual increase in the nuclear refractive index during the apoptosis process is observed, revealing the increase in molecular density and the decrease in cell volume. The strategy used in this study not only contributes to the understanding of internal environmental variations during cellular apoptosis but also provides a new platform for non-fluorescent all-fibre devices to investigate cellular events and to promote new progress in fundamental cell biochemical engineering.

**Introduction**

The intracellular microenvironment involves vital physiological characteristics of various cellular compartments. The fabrication of nano-probes for subcellular measurement is important to fully characterize cellular function. Single-cell interrogation at the nanoscale is carried out mainly via optical approaches, such as the use of fluorescent dyes[1-3], quantum dots or nanoparticles[4-11], nano-fibre probes[12,13], single nanowire probes[14-19], and planar photonic crystal nano-waveguides[20], mechanical approaches, such as the use of atomic force microscopy probes[21-23] and nanobeam arrays[24], and electrochemical approaches, such as the utilization of single metal nanowires[25,26], pillars or tube electrodes[27-31]. Dyes and quantum-dot-based probes have attracted much attention in biology, attributable to their photoluminescence (PL), and achieved great success in measuring the temperature of cells and their organelles[32,33], as they can be easily phagocytosed by cells due to their ultrasmall dimensions and are quite robust against external disturbances. However, these techniques suffer from some limitations, such as fluorescence bleaching and background fluorescence interruption. Optical nano-fibres have become increasingly popular owing to their excellent optical waveguiding property, high flexibility and ease of integration[34] and thus have been applied for the detection of intracellular pH[14] and ionic concentrations[15]. Notably, the existing fibre probes are passive in nature and serve only as conduits to guide light signals into/from a cell but not as a complete optical functional module for physical or chemical label-free sensing. Accordingly, most nanoprobe techniques still have to modify other nanoparticles, resulting in the low functional integration of structures.

The miniaturization of optical modules based on fibres is continuously pursued by researchers. Limited by manufacturing techniques, the scale of all-fibre functional modules is on the order of tens or hundreds of micrometers[35,36], which is not applicable compared to the sizes of most common biological cells. Fortunately, one-dimensional nanowires compensate for this deficiency. Their smaller dimensions and high robustness make them suitable for optical resonators[37], couplers[38], gratings[39], and lasers[40,41]. However, most of the current optical nano-devices are fixed on substrates, lacking portability and the plug-and-play feature. Moreover, local far-field incident light also implies low energy efficiency. For an indeed all-fibre label-free device, the excitation, transmission, detection and collection of light signals should be integrated in the same device without any chemical modification. Only in this way can the integration of the device, the convenience of manipulation and the efficiency of sensing be greatly improved. Under these circumstances, the combination of nanowire optical devices and fibre probes provides a novel configuration. Adhered to the tip of a tapered fibre, the nano-device can be freely moved, can be flexibly directed to anywhere and can safely penetrate the plasma membrane and enter biological cells, which is potentially useful in single-cell interrogation with high resolution and efficiency.

In this paper, we develop a compact all-fibre label-free nano-sensor based on a complete functional module composed of a fibre taper integrated with zinc oxide (ZnO) nano-gratings for the early monitoring of apoptosis in living cells, as illustrated in Fig. 1a. Cellular apoptosis, which is involved in cell turnover in many healthy adult tissues, is responsible for the focal elimination of cells during normal embryonic development. The visualization and monitoring of early apoptosis and the real-time tracking of cell status with high sensitivity in living cells are highly desired. We choose a single ZnO nanowire not only because of its low toxicity and high biological compatibility but also because it can serve as a waveguiding matrix and has low transmission loss and a relatively high refractive index (RI, n = 1.9831 at 655 nm), which are required for strong light confinement in the nanowire, especially in aqueous environments (n = 1.3310 at 655 nm). Moreover, the uniform diameter of the nanowire results in less physical damage to cellular structures and functions than caused by a conical optical fibre probe when penetrating a cell. With optical nano-gratings etched on the ZnO nanowire, the integrated fibre sensor is functionalized with the capability of simultaneously sending and receiving optical signals, being sensitive to changes in the surrounding environment, and can realize rapid, accurate and real-time sensing by being physically located in a single living cell. In our study, we find that the RI of the nucleus is higher than that of the cytoplasm in HeLa cells. More importantly, a gradual increase in nuclear RI during the apoptosis process after the invasion of hydrogen peroxide is observed, which illustrates an increase in molecular density and a decrease in cell volume[42]. Overall, the quantitative probing and analysis of the RI naturally present in single living cells during apoptosis may offer insight into the life activities within cells. Additionally, our strategy creates a precedent that allows a passive all-fibre nano-device for *in situ*, label-free, dynamic intracellular monitoring and may provide new ideas for a number of novel studies combining photonics with biochemistry.

**Results**

*Architectures of the nano-sensor*

The nano-sensor was fabricated by bonding ZnO nano-gratings on the tip of an optical fibre taper (Methods). The manufactured fibre taper had a conical transition region generally several millimetres in length and a long pigtail to allow prompt linking to other optical fibre components. Light travelling along the fibre can be effectively coupled into the nano-gratings, where part of the light is reflected back. Predictably, thinner nanowires cause fewer injuries to cell membranes during the puncture process; however, a larger diameter is more desirable in terms of light confinement and mechanical robustness. Therefore, we prioritized nanowires with diameters of 500-800 nm. The microscope image (Fig. 1c) of a typical probe shows that a single ZnO nanowire with gratings in the visible band (~660 nm) etched on the front end was successfully attached onto the tip of the fibre taper. The scanning electron microscopy (SEM) micrograph of the nanostructured probe (Fig. 1b) shows that the gratings on the ZnO nanowire included 40 shallow grooves with a period $\Lambda$ = 169 nm, resulting in a total length less than 7 μm, which is two orders of magnitude shorter than gratings fabricated in conventional optical fibres. Each groove was ~100 nm deep and ~80 nm wide and located at a position with a local radius of ~400 nm.

For a highly efficient nanowire probe, the optical output is closely confined to the ZnO nanowire, offering highly localized illumination. Owing to the excellent waveguiding properties of both the optical fibre and ZnO nanowire, the optical transmission loss is generally concentrated in the coupling region between them. As illustrated in our previous research[43], the normalized optical power flow coupled into the nanowire depends on the propagating constants of the two waveguides, which are related to the diameters of the two waveguides in the coupling region. To improve the coupling efficiency, we designed and fabricated probes with appropriate sizes. Figure 1b clearly shows that most light transmitted in the fibre probe is coupled into the nanowire. Benefiting from the high RI of ZnO, the external environmental RI hardly affects the optical coupling and light propagation.

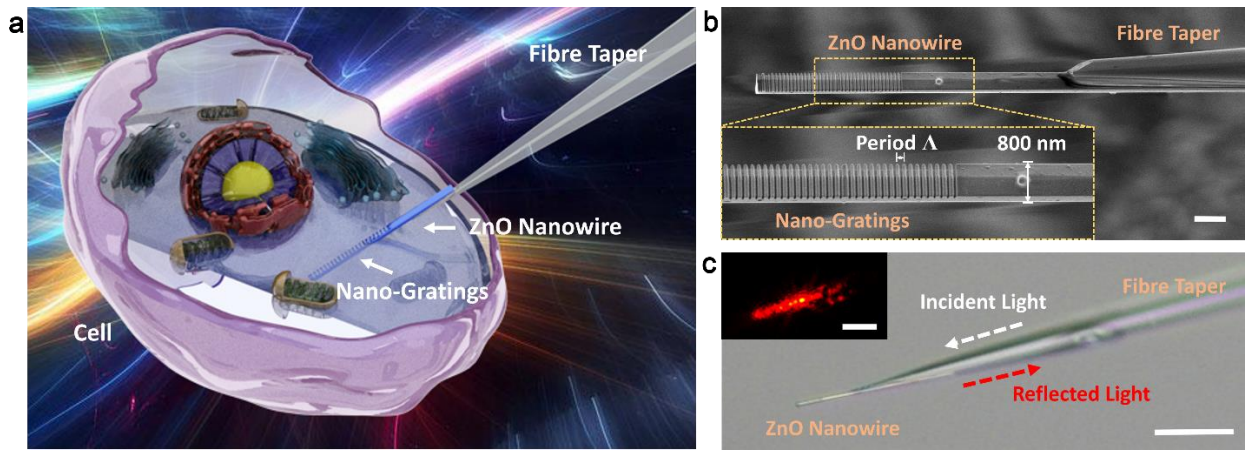

**Fig. 1. Schematic illustration and design of the all-fibre label-free nano-sensor. a** Schematic diagram of the ZnO nano-gratings integrated on a fibre bioprobe. **b** SEM image of the fibre sensor, with a nanowire diameter of 800 nm. Scale bar, 1 μm. Inset, details of the nano-gratings in the yellow dashed rectangle. **c** Microscope image of the fibre sensor with the nano-gratings against a bright field. Scale bar, 10 μm. Inset, Microscope image of the nanowire against a dark field, with incident light of 633 nm. Scale bar, 5 μm.

*Optical characteristics and calibration* **in vitro**

The setup used to test the all-fibre nano-sensor is illustrated in Supplementary Fig. 2. Figure 2a shows the typical reflectance spectrum of the fibre probe immersed in deionized water (Methods). To investigate the function of the ZnO nano-gratings, we measured the reflectance spectra of the fibre taper without (black curve) and with (blue curve) the ZnO nano-gratings. It is evident that the sole fibre taper reflects little light, while by contrast, the ZnO nano-gratings have a characteristic reflection wavelength at ~658 nm. The ratio of reflectance is approximately 30% compared to the end face of a sliced fibre. The extinction ratio, which indicates the ratio of the maximum to minimum reflection intensities, is ~11 dB, and the full width at half maximum (FWHM) is ~8 nm.

    A distinctive feature of optical gratings is that they are very sensitive to the RI of the external environment. With the change in RI, we discovered an apparent redshift in the peak reflection wavelength (Fig. 2b and Methods). Figure 2c (black line) demonstrates the linear relationship between the peak reflection wavelength of the nano-sensor and the surrounding RI, with a sensitivity of approximately 83 nm/RIU. The experimental result is in good agreement with the theoretical simulation (Supplementary Fig. 1). After calibration, the RI of pure phosphate buffered saline (PBS, HyClone) was measured to be 1.3345, again similar to previously reported data[44].

Using the shift in the spectrum is a common method in optical sensing, whereas the increase in temperature resulting from the high light power cannot be neglected, as it may cause predictable thermal damage to the cells. To make the experimental conditions more rigorous, a low-power laser at 655 nm was used as the light source to conduct single-cell sensing. Figure 2c (red line) shows the reflection power of 655 nm derived from Fig. 2b (red dashed line) with increasing RI. The linear relationship indicates that measuring the power change of a single wavelength is another feasible optical sensing approach. As illustrated in Fig. 3a, the laser (~800 nW) was injected into the fibre nano-sensor, and the reflected signal was collected with a power metre (Methods). The optical power passing through the fibre taper was hundreds of nanowatts, but the light was not completely incident into the cell because nearly half the light was lost in the coupling region between the fibre taper and the nanowire. With such low optical power, a possible temperature increase hardly occurred. The calibration indicates that with increasing RI, the reflection power at 655 nm will gradually decrease within a certain range (Fig. 2d, black line). Therefore, the RI information can also be obtained by measuring the light power at a single wavelength, with the experimental sensitivity of the nano-sensor being ~1280 %/RIU for an RI ranging from 1.330 to 1.350 (red line). This result is generally consistent with the experimental results presented in Fig. 2c (red line).

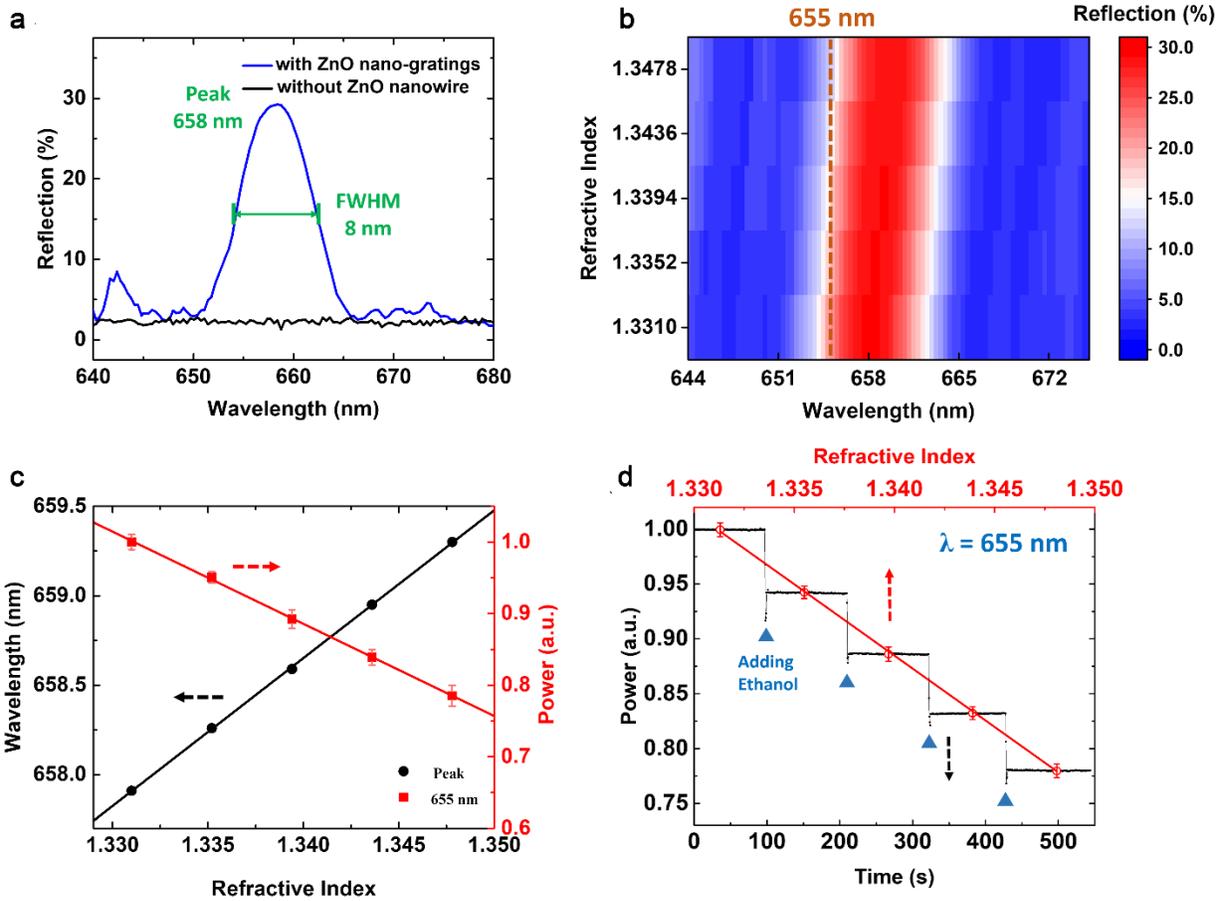

**Fig. 2. Optical characteristics and RI sensitivity of the nano-sensor. a** Reflectance spectra of the fibre probe with (blue line) and without (black line) ZnO nano-gratings. The peak wavelength is 658 nm. **b** Colour map of spectra with different external RIs. **c** Shift in the peak reflection wavelength (black line) and the change in power at 655 nm (red line) as a function of the external RI. The red line is derived from the values in **b**. **d** Dynamic response of the reflection power at 655 nm with the addition of ethanol (black line) and the power at 655 nm as a function of the external RI (red line). The figure closely coincides with the red line in **c**.

*Safe penetration tests*

To demonstrate that the nanowire probes incurred negligible cell membrane damage during insertion and extraction, fluorescent dyes were used to verify whether the cell membranes were ruptured (Methods). Optical fibre probes with and without a ZnO nanowire were manipulated to penetrate HeLa cells. Figure 3b shows the tested cell (red circle) after insertion of a fibre probe with an ~500 nm conical tip. The PL intensity decreased significantly compared with that of other cells 3 minutes after puncture, which indicated that the cell membrane was likely to be damaged, resulting in the outflow of dyes. Figure 3c shows the tested cell (red circle) after insertion of the ZnO nanowire-integrated fibre probe with an ~800 nm tip. Owing to the uniform diameter of the nanowire, the PL intensity of the tested cell showed no apparent differences compared with the surrounding cells within several minutes after puncture, indicating that the cell membrane was still relatively intact. The observation of fluorescence from the cells shows that the cells continued to hydrolyse fluorescein diacetate to generate fluorescein inside the cells. Therefore, it is feasible to use the ZnO nanowire-integrated fibre sensor to implement intracellular measurements.

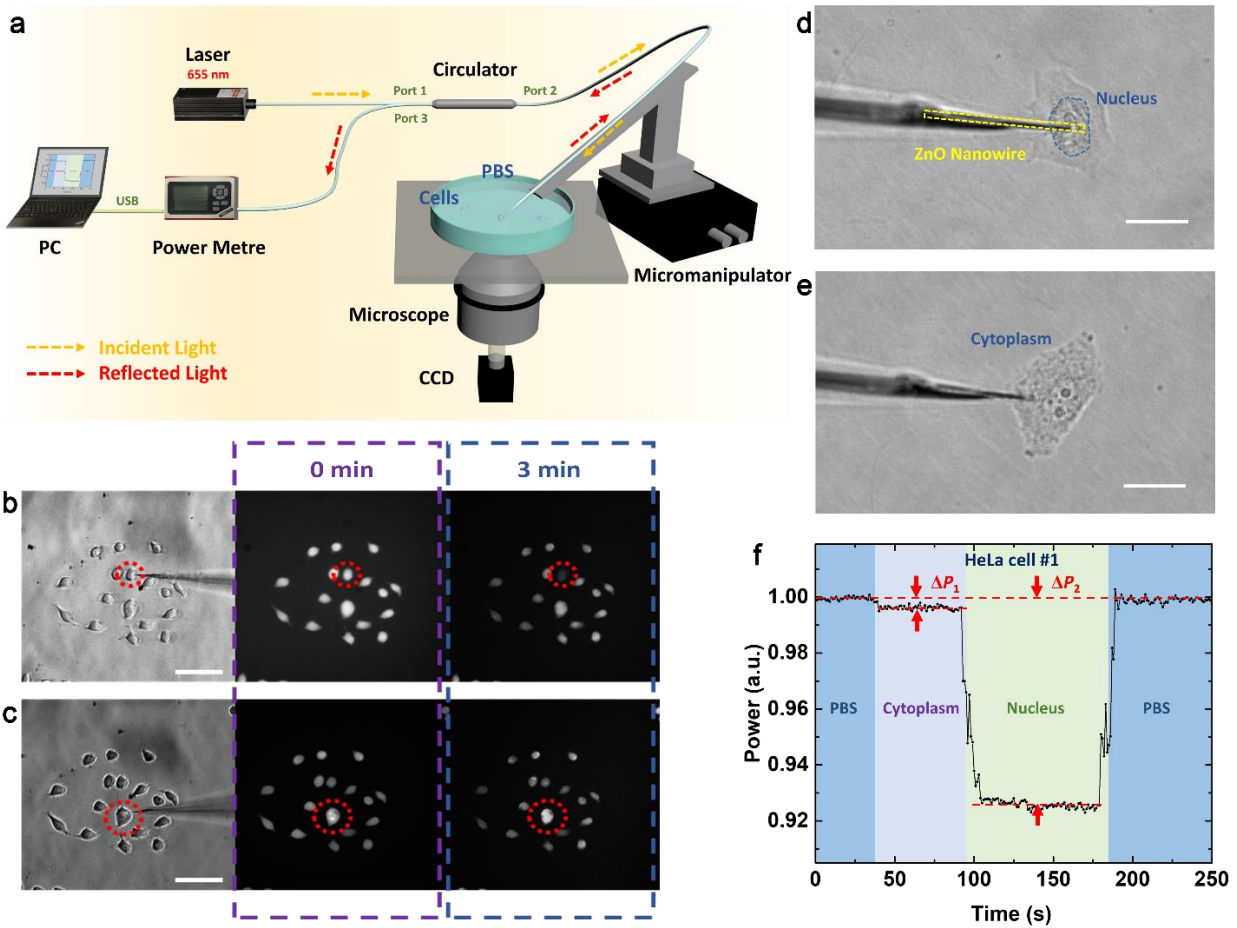

**Fig. 3. Intracellular calibration and measurements of the nano-sensor. a** Schematic diagram of the experimental setup used to calibrate the RI sensitivity of the probe and measure the intracellular RI of HeLa cells. A low-power laser at 655 nm (~800 nW) was injected into the pigtail of the fibre probe through port 2 of a visible circulator. The output light reflected at the nano-gratings was transmitted through port 3 of the circulator, which was connected to an optical power metre to record the reflection power. **b**, **c** Microscope and PL images of the tested cell (in red circle) and other reference cells. The tested cell was penetrated by a fibre probe with an ~500 nm conical tip (**b**) and a ZnO nano-grating-integrated fibre probe with an ~800 nm tip (**c**). Scale bar, 40 μm. **d**, **e** Microscope images of the ZnO nanowire-integrated fibre probe inserted into the nucleus (**d**) and cytoplasm (**e**) of single HeLa cells. Scale bar, 10 μm. **f** Normalized reflection power in the cytoplasm and nucleus of a single HeLa cell, indicating the difference in RI.

*Intracellular experiments on RI sensing*

Figure 3d, e show that the nucleus and cytoplasm regions of single HeLa cells are clearly visible in the microscope images even after insertion of the ZnO nano-grating-integrated fibre sensor (Methods). The nanowire is robust against cell penetration, with no obvious mechanical deformation. As illustrated in Fig. 3f, the normalized reflection power did not decline significantly when the probe was transferred from PBS into the cytoplasm. However, it decreased by ~7% when the probe entered the nucleus. Notably, the experiment was carried out in a constant temperature environment to eliminate the influence of temperature fluctuations. We tested four single HeLa cells, and the experimental results in Table 1 and Supplementary Fig. 3 indicate that the mean RI of the cytoplasm and the nucleus in the HeLa cells was 1.3348 and 1.3402, respectively. Since the RI is generally proportional to the molecular concentration, which is related to the viscosity of solutions, the molecular density of the nucleus is deduced to be higher than that of the cytoplasm. The relatively stable fluctuation of the reflection power also indicated that the signal light had little thermal effect on the cells. Previous studies using optical diffraction tomography also drew similar

conclusions[42]. Compared with traditional far-field detection techniques, our method is more energy efficient, as only very low-power probe light is required.

*Intracellular dynamic monitoring of cellular apoptosis*

Information on RI changes in the nucleus when the cell is gradually apoptotic under external stimuli is of great importance for research on cell life events. The advantage of using nano-gratings instead of fluorescence for single-cell detection is that the nano-configuration is more stable and reliable for long-term monitoring. After the probe punctured the nucleus of a HeLa cell, 10 μL of 20 mM $H_2O_2$ PBS solution was injected into the medium. Hydrogen peroxide (n =1.3350 at 655 nm) is a strong oxidant that breaks down cell membranes and is thus fatal to cells. The microscope images of the detected HeLa cells (Fig. 4a-d) show the variation in cell appearance. Simultaneously, a reduction in the reflection intensities was observed in the nucleus, indicating a continuous increase in RI with time (Fig. 4e). We tested three single HeLa cells, and the mean nuclear RI after cell apoptosis was 1.3442, as indicated in Table 2 and Supplementary Fig. 5. The numbers of pixels that the cell nucleus occupied in the microscope images (Fig. 4a-d) during the process of cell apoptosis were counted, which could roughly reflect the volume of the cell nucleus (see Fig. 4e). The volume of the cell nucleus gradually contracted, leading to an increase in molecular density, which is one of the possible reasons for the increase in the nuclear RI.

A mature fluorescence technique was used to examine the cell viability of the tested cells and other normal cells during the process of apoptosis (Methods). The fluorescent dye we chose presents little toxicity to cells, can infiltrate through cell membranes, and is often used for nuclear staining and cell apoptosis tests (Supplementary Fig. 4). As shown in Fig. 4e, the intensity of the PL signal underwent an apparent increase 40 minutes after $H_2O_2$ was injected, confirming that the cells had indeed been apoptotic. Moreover, the PL intensity of the tested cell (yellow dashed circle) had no obvious difference from that of other surrounding cells, which again indicated that the ZnO nano-gratings had little influence on the cell life states.

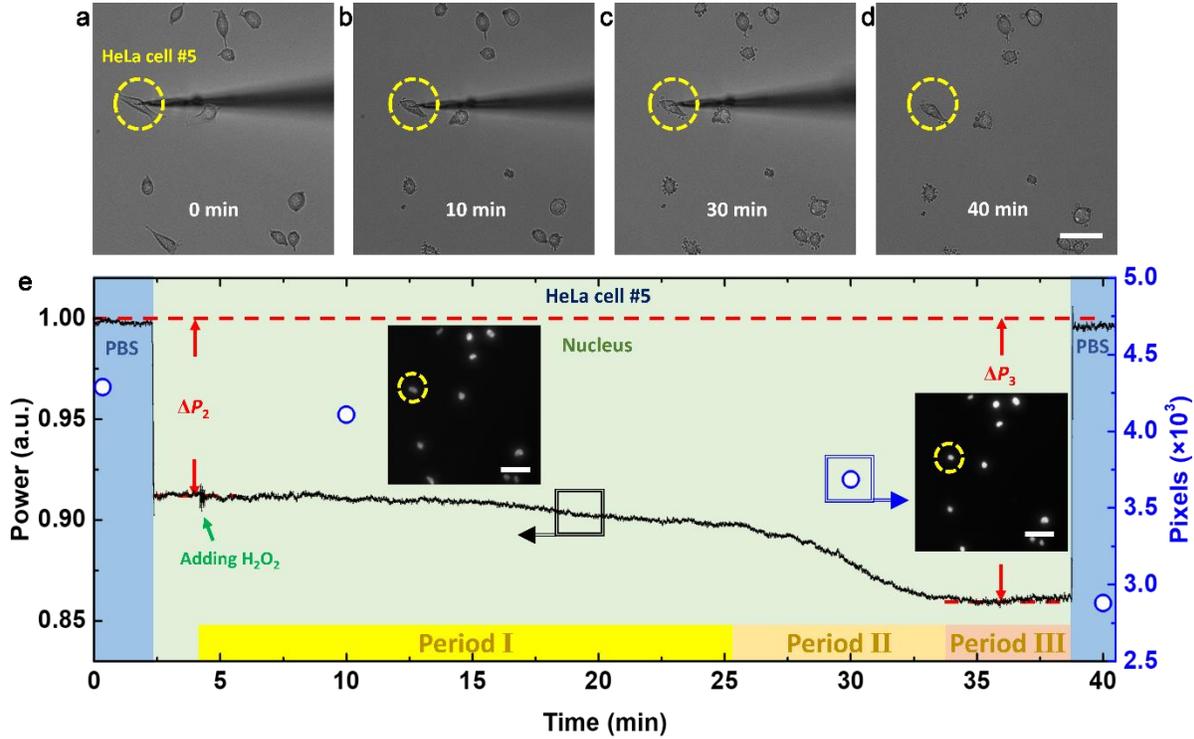

**Fig. 4. Real-time long-term early monitoring of HeLa cell apoptosis. a-d**, Microscope images of single tested HeLa cells (yellow dashed circle) and other reference cells during the apoptosis process. **a** Immediately after the addition of $H_2O_2$. **b** 10 min after the addition of $H_2O_2$. **c** 30 min after the addition of $H_2O_2$. **d** Apoptotic cells 40 min after the addition of $H_2O_2$. Scale bar, 20 μm. **e** Normalized reflection power in the nucleus of a single tested HeLa cell during apoptosis under the stimulation of $H_2O_2$ (black line), indicating a continuous increase in the nuclear RI. The numbers of pixels that the cell nucleus occupies in the microscope images (blue circle) roughly indicate the volume of the nucleus. Inset, PL images of living and apoptotic HeLa cells. Scale bar, 20 μm.

**Discussion**

The gradual increase in nuclear RI went through three periods. During the first period, which lasted

~20 minutes, the process of apoptosis was slow. The cytoplasm was gradually dissolved under the corrosion of hydrogen peroxide, but the morphology of the nucleus did not have a notable impact (Fig. 4b), and the RI of the nucleus increased slowly. During the second period, which lasted ~10 minutes, large areas of the cytoplasm dissolved, and the nucleus began to shrink (Fig. 4c). The nuclear RI rapidly increased by ~0.0040 with decreasing nuclear volume. During the last period, the cells had little vitality, and the nuclear RI tended to become stable. Although the exact reasons for the changes in the RI of the nucleus are still unknown, our results imply that the shape and volume of the cell are at least partly responsible.

In conclusion, we developed a versatile, biocompatible, portable and reusable ZnO nano-grating-integrated all-fibre label-free nano-sensor that enables *in situ* early monitoring of cellular apoptosis. The nano-device integrates signal transmission, sensing and collection, with high integration, operability and sensitivity. By inserting the nanoscale functional optical module into single living HeLa cells, we can detect different features of organelles. The highly flexible and robust nano-sensor allowed us to reveal that the mean RI of the nucleus is 0.0054 higher than that of the cytoplasm in single HeLa cells, which to some extent verifies the higher molecular density of the nucleus than that of cytoplasm. Moreover, the stable and nondegradable configuration is promising for the long-term monitoring of cellular environmental variations, which are universally present in all cell life activities. We observed a decrease in the nuclear volume and detected the nuclear RI during cell apoptosis. The results show that the final mean RI is 0.0040 higher than that of normal living cells, linking the RI and the volume of the nucleus to the state of living cells. This study is the first to realize apoptosis sensing and tracking with an all-fibre label-free nano-device in living cells. Real-time probing of cellular apoptosis will not only attract more attention to cell life events but also broaden the application scope of all-fibre nano-sensors.

**Methods**

*Nano-sensor fabrication*

The fibre taper, with a sharp tip whose diameter is hundreds of nanometres, is fabricated from a single-mode optical fibre (NUFERN, 630-HP) drawn with a commercial pipette puller (Sutter Instrument Co., MODEL P2000). To obtain a sharp and smooth taper profile, we optimized the $CO_2$ laser power and pulling velocity. ZnO nanowires were synthesized using the chemical vapour deposition (CVD) method, with lengths ranging from dozens to hundreds of microns and diameters less than 1,000 nm. A single ZnO nanowire was first picked up from the glass slide using a fibre probe immobilized on a high-precision 3D translational stage (Thorlabs) and thereafter transferred along the tip of the fibre taper under a microscope (Nikon, L-1M). A UV-curable adhesive (EFiRON UVF PC-375, Luvantix) was chosen to firmly fix the nanowire to improve the mechanical stability of the probe, which is a fundamental requirement for use in intracellular sensing. A focused ion beam (FIB, FEI, Helios 600i) was used to etch the nano-gratings at visible wavelengths (~660 nm) on the front end of the nanowire. The shape of the gallium ion beam was cylindrically symmetric, with a diameter of ~10 nm. With this accuracy, the groove size could be precisely controlled.

*Setup for spectral characterization* **in vitro**

The fibre nano-sensor was immobilized on the high-precision 3D translational stage (Thorlabs), and the nano-gratings were immersed in deionized water. The process was carried out under a microscope with a charge-coupled device (CCD, Nikon) camera. A supercontinuum source (Wuhan Yangtze Soton Laser Co., Ltd, SC-5) was utilized to characterize the reflection spectrum of the fibre probe. The signal light was injected into the pigtail of the fibre probe through a visible circulator (Thorlabs, 670 nm). Then, the output light reflected at the nano-gratings was transmitted through the circulator to an ultraviolet-visible spectrometer (Ocean Optics) to record the spectrum, which ranged from 640 to 680 nm.

*Setup for power characterization* **in vitro**

The fibre nano-sensor was immobilized on the high-precision 3D translational stage (Thorlabs), and the tip was immersed in deionized water. A low-power laser (Changchun New Industries Optoelectronics Tech Co., Ltd.) at 655 nm (~800 nW) was injected into the pigtail of the fibre probe through a visible circulator (Thorlabs, 670 nm). Then, the output light reflected at the nano-gratings was transmitted through the circulator to an optical power metre (OPM, Thorlabs, PM100D) to record the reflection power.

*Methods for RI calibration of the nano-sensor*

A pipette was used to inject ethanol (n = 1.3599 at 655 nm) into 1 mL of deionized water, and the output spectrum/power was detected. Every time 300 μL of ethanol was injected, we discovered a redshift in the peak reflection wavelength during spectral characterization and a reduction in the reflection power at 655 nm during power characterization. We added ethanol several times, and the resulting RIs were calculated based on the concentration of ethanol in the solution.

*Insertion of the nano-sensor*

The ZnO nanowire-integrated fibre sensor was fixed on an electrically controlled 3D

micromanipulator (DWCQ-2), and only the ZnO nano-grating region was slowly inserted into the cytoplasm and nucleus of single HeLa cells under a microscope (Olympus, IX51).

*Preparation for safe penetration tests*

HeLa cells were stained with 10 μg/mL fluorescein diacetate (Sigma, F7378-5G) and incubated for 10 minutes. The dye presents little toxicity to cells and can be phagocytosed into the cytoplasm. A mercury lamp (Olympus, U-RFL-T) was utilized to excite the PL of the fluorescent dyes.

*Preparation for fluorescence cytoactive detection*

HeLa cells were stained with 10 μg/mL Hoechst 33342 (100X) dye (Beyotime) and incubated in the dark for 10 minutes. A mercury lamp (Olympus, U-RFL-T) was utilized to excite the PL of the fluorescent dyes. The PL signal ($\lambda = 460$ nm) passed through an emission filter to remove light originating from the excitation ($\lambda = 346$ nm) and detection ($\lambda = 655$ nm) sources and was finally collected by the CCD.

**Data availability**

Source data for all figures are available from the corresponding author.

**Code availability**

The code that supports simulated results of this study is available from the corresponding authors upon reasonable request.

**Acknowledgements**

This research was sponsored by the National Natural Science Foundation of China (61925502, 61535005, 22025403) and the National Key R&D Program of China (2017YFA0303700).

**Author contributions**

F. X. and D. J. developed the concept and conceived the experiments. D. L. fabricated and assembled the nano-sensor, performed the experiments, and wrote the original draft manuscript. N. W., T. Z. and W. Z. cultivated HeLa cells, and N.W. assisted in completing the cell experiments. G. W. and Y. T. performed theoretical simulation and analysis on the structure of the sensor. Y. X. provided Figure 1a. Q. D., S.G. and J. Y. fabricated ZnO nanowires. Y. L. advised on the optical measurement. F. X., D. J. and D. L. analyzed the data, and reviewed and edited the paper. All authors contributed to discussions and writing of the manuscript. F. X. supervised the research.

**Competing interests**

The authors have no conflicts of interest to declare.

**Figure legends**

**Fig. 1. Schematic illustration and design of the all-fibre label-free nano-sensor. a** Schematic diagram of the ZnO nano-gratings integrated on a fibre bioprobe. **b** SEM image of the fibre sensor, with a nanowire diameter of 800 nm. Scale bar, 1 μm. Inset, details of the nano-gratings in the yellow dashed rectangle. **c** Microscope image of the fibre sensor with the nano-gratings against a bright field. Scale bar, 10 μm. Inset, Microscope image of the nanowire against a dark field, with incident light of 633 nm. Scale bar, 5 μm.

**Fig. 2. Optical characteristics and RI sensitivity of the nano-sensor. a** Reflectance spectra of the fibre probe with (blue line) and without (black line) ZnO nano-gratings. The peak wavelength is 658 nm. **b** Colour map of spectra with different external RIs. **c** Shift in the peak reflection wavelength (black line) and the change in power at 655 nm (red line) as a function of the external RI. The red line is derived from the values in **b**. **d** Dynamic response of the reflection power at 655 nm with the addition of ethanol (black line) and the power at 655 nm as a function of the external RI (red line). The figure closely coincides with the red line in **c**.

**Fig. 3. Intracellular calibration and measurements of the nano-sensor. a** Schematic diagram of the experimental setup used to calibrate the RI sensitivity of the probe and measure the intracellular RI of HeLa cells. A low-power laser at 655 nm (~800 nW) was injected into the pigtail of the fibre probe through port 2 of a visible circulator. The output light reflected at the nano-gratings was transmitted through port 3 of the circulator, which was connected to an optical power metre to record the reflection power. **b**, **c** Microscope and PL images of the tested cell (in red circle) and other reference cells. The tested cell was penetrated by a fibre probe with an ~500 nm conical tip (**b**) and a ZnO nano-grating-integrated fibre probe with an ~800 nm tip (**c**). Scale bar, 40 μm. **d**, **e** Microscope images of the ZnO nanowire-integrated fibre probe inserted into the nucleus (**d**)

and cytoplasm (**e**) of single HeLa cells. Scale bar, 10 μm. **f** Normalized reflection power in the cytoplasm and nucleus of a single HeLa cell, indicating the difference in RI.

**Fig. 4. Real-time long-term early monitoring of HeLa cell apoptosis. a-d**, Microscope images of single tested HeLa cells (yellow dashed circle) and other reference cells during the apoptosis process. **a** Immediately after the addition of $H_2O_2$. **b** 10 min after the addition of $H_2O_2$. **c** 30 min after the addition of $H_2O_2$. **d** Apoptotic cells 40 min after the addition of $H_2O_2$. Scale bar, 20 μm. **e** Normalized reflection power in the nucleus of a single tested HeLa cell during apoptosis under the stimulation of $H_2O_2$ (black line), indicating a continuous increase in the nuclear RI. The numbers of pixels that the cell nucleus occupies in the microscope images (blue circle) roughly indicate the volume of the nucleus. Inset, PL images of living and apoptotic HeLa cells. Scale bar, 20 μm.

**Tables**

**Table 1.** RI measured in the cytoplasm and nucleus of 4 HeLa cells.

| Cell no. | Decrease in power ($\Delta P_1$) in cytoplasm (%) | RI in cytoplasm | Decrease in power ($\Delta P_2$) in nucleus (%) | RI in nucleus |
|---|---|---|---|---|
| 1 | 0.34 | 1.3348 | 7.41 | 1.3403 |
| 2 | 0.69 | 1.3350 | 9.59 | 1.3420 |
| 3 | 0.13 | 1.3346 | 5.62 | 1.3389 |
| 4 | 0.25 | 1.3347 | 6.39 | 1.3395 |

**Table 2.** Measured RI of the nucleus in 3 apoptotic HeLa cells.

| Cell no. | Decrease in power ($\Delta P_2$) in nucleus before apoptosis (%) | Nuclear RI before apoptosis | Decrease in power ($\Delta P_3$) in nucleus after apoptosis (%) | Nuclear RI after apoptosis |
|---|---|---|---|---|
| 5 | 8.16 | 1.3408 | 13.89 | 1.3453 |
| 6 | 6.92 | 1.3399 | 11.53 | 1.3435 |
| 7 | 7.59 | 1.3405 | 11.92 | 1.3438 |

# Figures

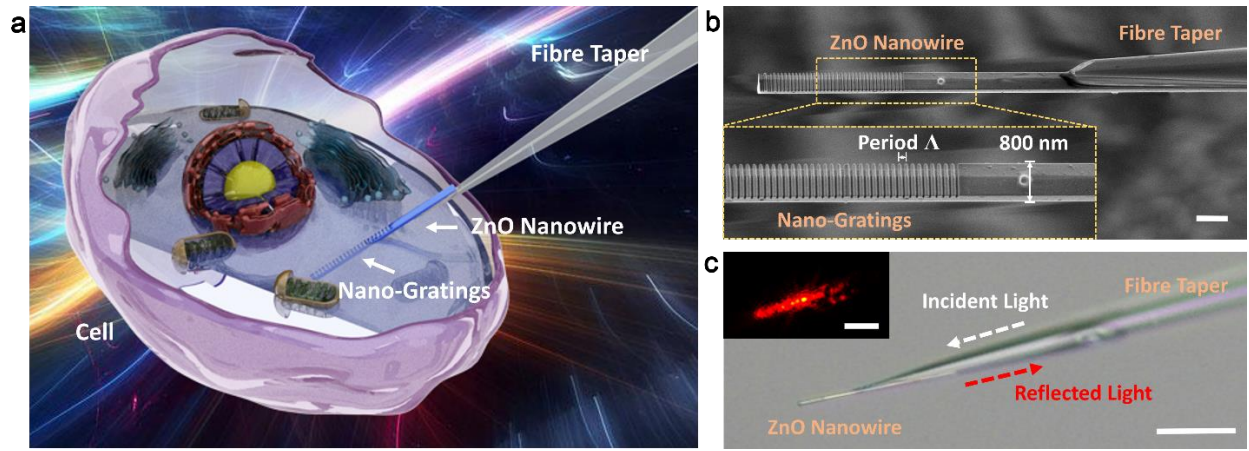

**Fig. 1**

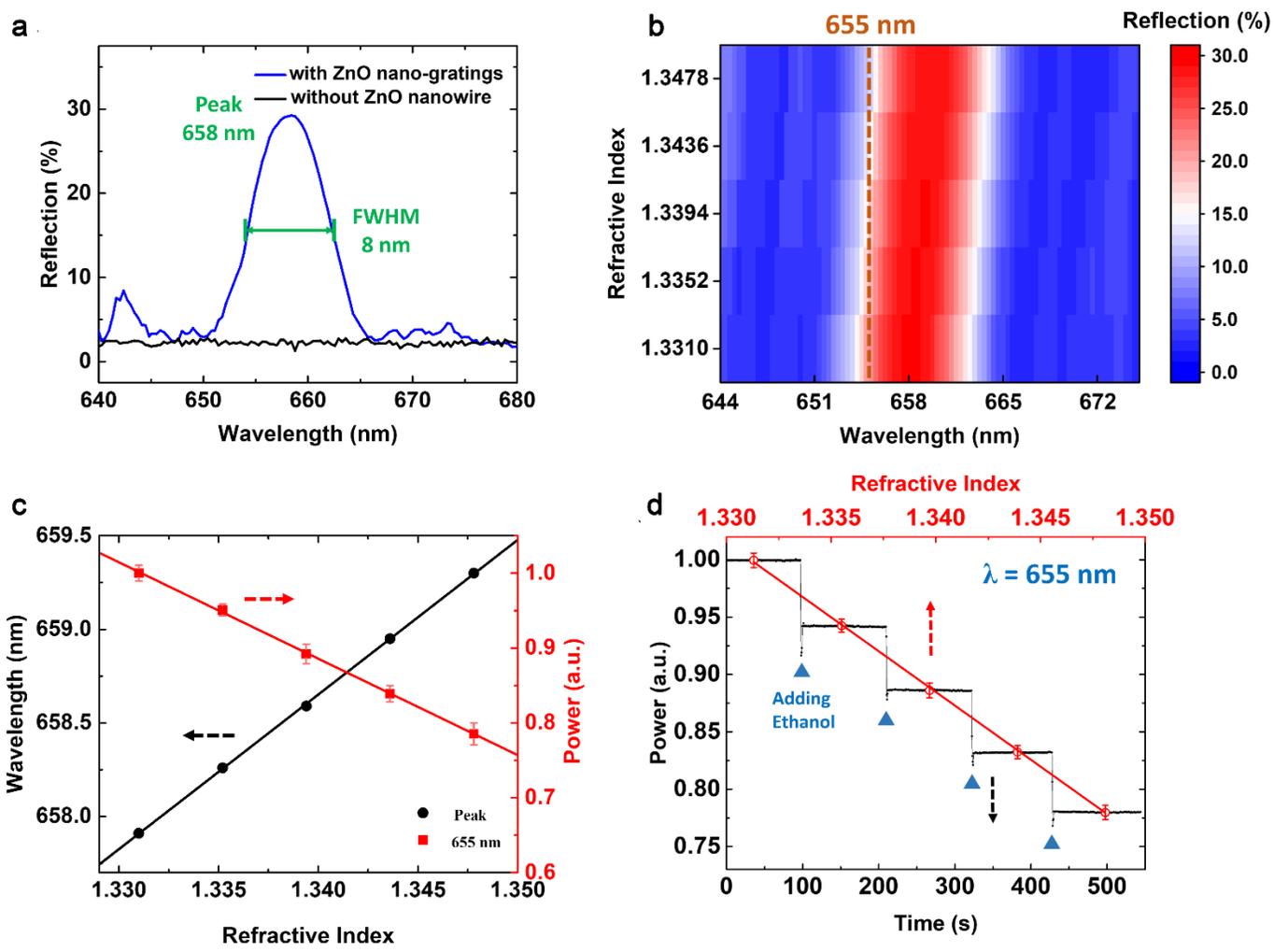

**Fig. 2**

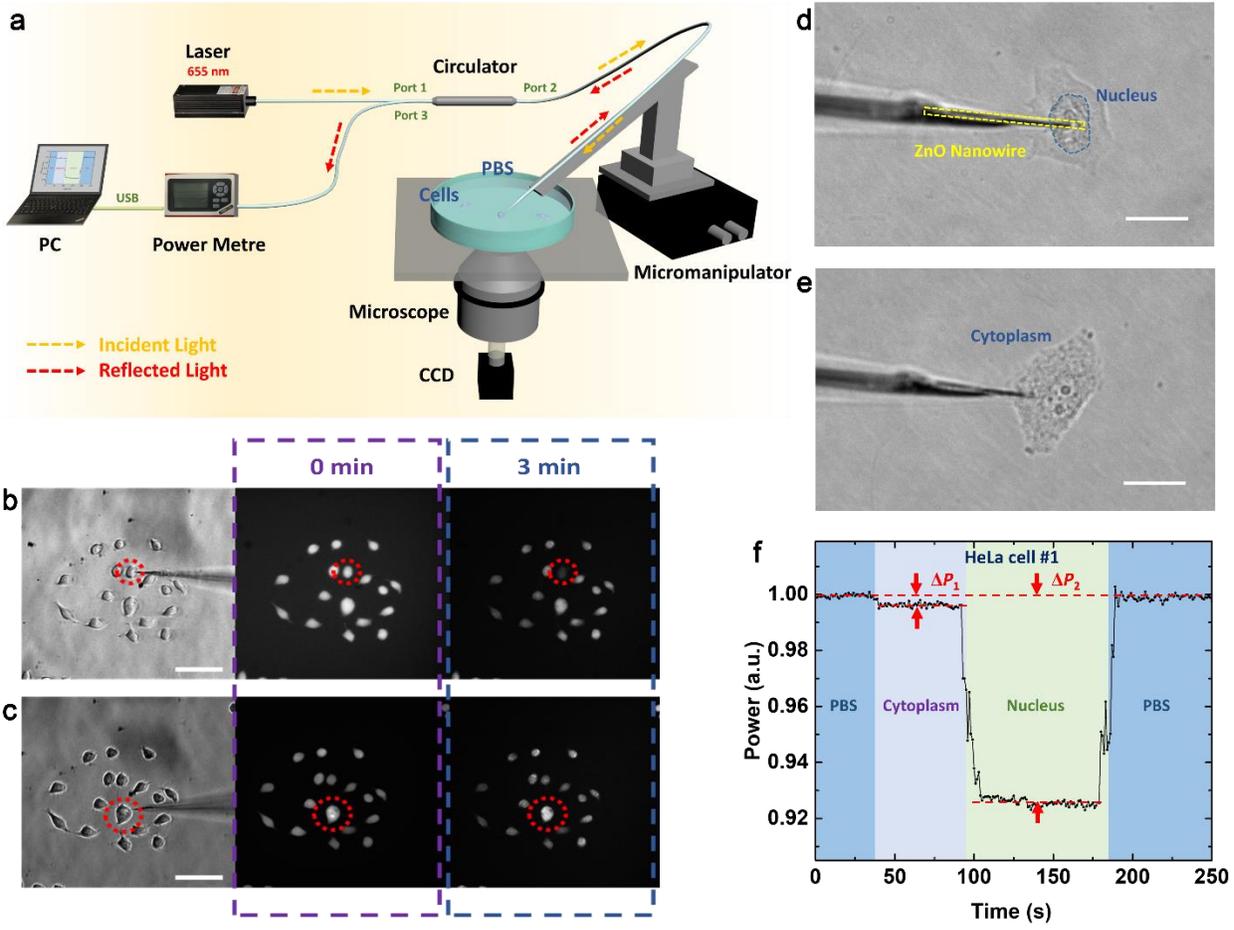

**Fig. 3**

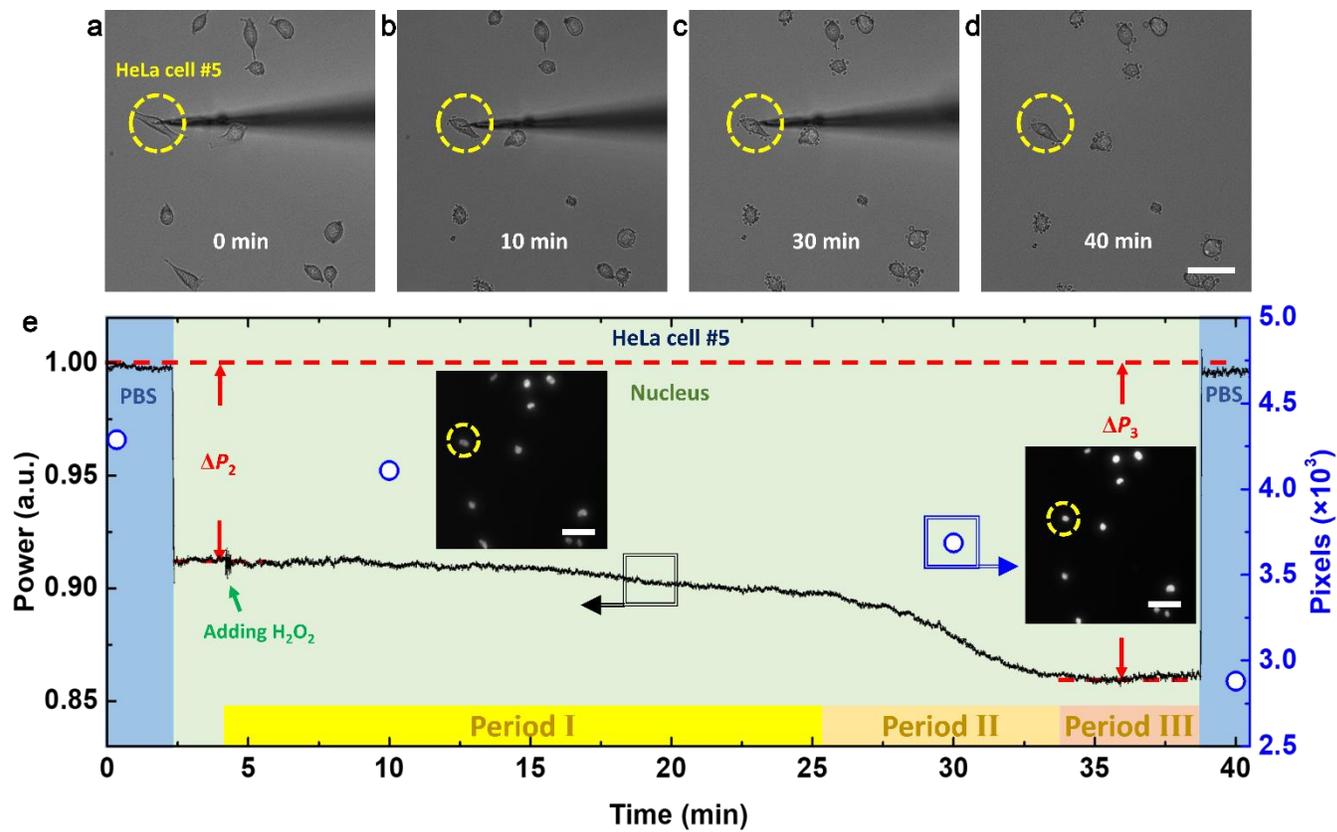

**Fig. 4**